\documentclass[prd,aps,twocolumn,showpacs,floatfix]{revtex4-2}

\usepackage{dcolumn}
\usepackage{amsmath}
\usepackage{amssymb}
\usepackage{bm}
\usepackage{graphicx,subfigure}
\usepackage{epsf,epsfig}
\usepackage{hyperref}
\usepackage{color}
\usepackage{enumerate}

\newcommand{\bq}{\begin{equation}}
\newcommand{\eq}{\end{equation}}
\newcommand{\bqn}{\begin{eqnarray}}
\newcommand{\eqn}{\end{eqnarray}}
\newcommand{\nb}{\nonumber}
\newcommand{\lb}{\label}
\newcommand{\IG}{_{_{\text{Simple}}}}

\begin{document}

\title{Non-local Interactions are Essential Elements for Dark Matter Halo Stability: A Cross-Model Study}
\author{Ahmad Borzou$^{1,  2}$}
\email[]{ahmad\_borzou@baylor.edu}
\affiliation{$^1$ EUCOS-CASPER, Department of Physics, Baylor University,
Waco, TX 76798, USA,
$^2$ CompuFlair,  Houston,  TX 77064,  USA
}
\date{\today}
\begin{abstract}
This paper introduces a comprehensive methodology for examining the stability of dark matter (DM) halos, emphasizing the necessity for non-local inter-particle interactions, whether they are fundamental or effective in nature, to maintain halo stability. We highlight the inadequacy of vanilla cold collision-less DM models in forecasting a stable halo without considering a ``non-local'' interaction in the halo's effective free energy, which could potentially arise from factors like baryonic feedback, self-interactions, or the intrinsic quantum characteristics of dark particles. The stability prerequisite necessitates significant effective interactions between any two points within the halo, regardless of their distance from the center.  The methodology proposed herein offers a systematic framework to scrutinize the stability of various DM models and refine their parameter spaces. We deduce that DM halos within a model, where the deviation from the standard cold collision-less framework is confined to regions near the halo center, are unlikely to exhibit stability in their outer sectors.
In our study, we demonstrate that the issue of instability within DM halos cannot be addressed adequately using perturbative quantum effects.  This issue is less pronounced for fermionic DM but suffers from a higher degree of severity when considering bosonic DM. We find that halos made of bosons with notable quantum effects have sharp edges, while those made of fermions show more diffuse boundaries extending toward infinity.
To present the potentials of the cross-model approach,  we explore the broadest form of the effective free-energy around a chosen mass profile.  Next,  as a show case study,  we employ a model where the deviation from the standard cold collision-less DM model is represented by a two-body interaction in the effective free-energy to show how to use observations to investigate universal classes of DM models. 
\end{abstract}                                                                                   
\maketitle

\section{Introduction} 
The $\Lambda$-CDM cosmological model, which characterizes dark matter (DM) as a cold and collision-less gas interacting with visible matter solely via gravity, effectively accounts for several observed phenomena. These phenomena include the cosmic microwave background's (CMB) correlation function \cite{Planck:2018vyg}, supernova redshift surveys \cite{Pan-STARRS1:2017jku}, and the clustering of galaxies \cite{BOSS:2016wmc}. Furthermore, the model aligns with the observation that smaller structures develop earlier in cosmic history \cite{Planck:2018vyg, Chabanier:2019eai}. Apart from a recent discrepancy in the Hubble constant value \cite{DiValentino:2021izs}, the $\Lambda$-CDM model successfully interprets large cosmic scales.

DM simulations that start from early universe initial density perturbations predict the formation of DM halos with dense and steep profiles towards the center, creating a cusp. However, this prediction contradicts observations from rotation curves of stars near galactic centers and gases in the outskirts, as well as stellar velocity dispersion data. These observations suggest a more constant mass density, or a core, at the center of galaxies \cite{1994Natur.370..629M, KuziodeNaray:2007qi, Oh:2008ww, 2011ApJ...742...20W, Oh:2015xoa}.

This inconsistency between DM simulations and observational data can be reconciled by considering baryonic effects on dark matter. However, unlike DM N-body simulations, these effects are not fundamentally modeled. Instead, aspects like star formation and viscosity are directly integrated into the simulations using numerous free parameters, which are then adjusted to align with observations \cite{Rosswog2009AstrophysicalSP}. The wide parameter selection flexibility in the current versions is suboptimal, although incorporating visible matter effects remains critical.

While N-body simulations may explain observed DM mass profiles by incorporating baryonic feedback, their numerous free parameters potentially allow the explanation of non-physical phenomena as well. To evaluate N-body simulations and other alternate DM paradigms, it is necessary to examine their predictions against observations not used in parameter tuning. One such observation is the stability of halos.  

We aim to initiate a systematic investigation of the stability of DM halos in this article.  The stability of DM halos can be probed by predicting the position dependent halo stability in a given DM scenario and comparing it with observational data. Furthermore, studying the stability of DM halos can enable cross-model comparisons and classifications of DM scenarios into universal categories, which can facilitate a ``coarse-grained-assessment" of DM models in light of the current observational limitations and the abundance of theoretical possibilities.

A key requirement for long-term stability is the satisfaction of the Vlasov-Poisson equation, which ensures that the net force on DM in the halo is zero, leading to dynamical stability \cite{binney2011galactic, 2009arXiv0904.2443L}. However, fluctuations in an N-body system are inevitable. If the system is not at the minimum of effective free energy, these fluctuations can rapidly increase, destabilizing the halo. This means that dynamical stability does not necessarily guarantee ``thermodynamic" stability, and a dynamically stable halo could still experience gravothermal catastrophe or collapse \cite{Chavanis:2014woa, 2020EPJP..135..290C}. Nevertheless,  ``thermodynamic" stability does imply dynamic stability \cite{1980ApJ...238.1101I}.

To investigate ``thermodynamic" stability, Landau damping and violent relaxation should be considered to find a solution minimizing the N-body system's free energy.  While this approach is robust, it increases the complexity of the calculations and depends on the DM model.
Existing studies on fermionic DM ``thermodynamic" stability can be found in \cite{Bilic:1998mc, Roupas:2018zie, Alberti:2018lqk}.  For some DM models, more attention has been devoted to Vlasov-Poisson dynamical stability, with less focus on long-term stability.

Given the complexity and model-dependence of the current approach to ``thermodynamic" stability, this article employs Landau's field-theoretic method to investigate long-term stability states of halos. This approach avoids dealing with the specifics of the DM model, instead focusing on the collective system's symmetries. Intriguingly, a broad spectrum of seemingly different models fall under one universality class, complying with the same statistical equations determined almost exclusively by the symmetries of the N-body system, barring highly entangled quantum systems where topology plays a role. This approach allows us to investigate a wide array of DM models and their free parameters within a single study. The unique aspect of our approach is its capacity to transfer results between different DM models.

In this article, we demonstrate that no self-gravitating classical model of DM can predict a stable halo unless a ``non-local" interaction between mass densities is included in the effective free energy of the halo. This interaction could be collective, for example due to baryonic feedback or self-interactions, or resulting from the quantum nature of dark particles, whether fermionic or bosonic. This stability condition demands substantial interactions between any two locations in the halo, even if both are far from the center. Therefore, if a DM model's deviation from the standard cold, collision-less scenario is confined to regions near the halo's center, the halo will still be unstable. Consequently, models like the cold, collision-less DM with baryonic feedback, where visible matter is located at the center,  are unlikely to predict pressure-supported stable outer halo regions. 

To showcase the potential of the field theoretic approach to studying DM halos, we expand the most general form of effective free-energy around an arbitrary mass profile.  We then choose a model whose deviation from the standard cold, collision-less model is a two-body interaction in the effective free-energy.  We demonstrate that even with a ``non-local" interaction, the halo may still be unstable for certain forms of the interaction.  Moreover,  using the showcase,  
we demonstrate that fluctuations of DM mass density around an empirically determined mass profile,  are contingent on the universal class of DM scenarios. As such, it becomes possible to restrict their parameter space.  Importantly,  any reduction in the parameter space of a particular universal class extends to all DM models within that category.

The structure of the article is as follows.  In section \ref{Sec:Non-Interactive-DM}, we establish the effective free energy of a simple, cold, collision-less DM halo, demonstrating its inherent instability. In section \ref{Sec:Interactive-DM}, we introduce DM interactions into the effective free-energy equation, elucidating the necessity of non-local interactions to stabilize the halo.
Proceeding to section \ref{Sec:Quantum-DM}, we formulate the effective free-energy of a halo incorporating non-interactive DM quantum effects. The equivalence between this quantum model and a classic interactive DM model, in the context of effective free-energy, is demonstrated. Further, we explore a particular model where quantum effects can be analyzed via the perturbation method.
In section \ref{Sec:General-DM-perturbation}, we design the most encompassing model of DM perturbations, illustrating its potential to investigate universal classes of DM perturbation models. Finally, in section \ref{Sec:Conclusion}, we draw our conclusions.

\section{Non-Interactive Classic DM}
\lb{Sec:Non-Interactive-DM}
We begin this section by deriving the effective free energy functional of mass density for a halo of cold collision-less DM, neglecting the effects of baryons. We consider a small volume element $\Delta V$ at a position $x$ in the halo.  The number of particles in this volume element is $N(x) = \frac{\rho(x)}{m} \Delta V$, where $\rho(x)$ is the mass density and $m$ is the particle mass. We assume that the system is in a steady state, so that the probability of the volume's state is equal to
\bqn
{\cal{P}}(x) = \exp\left(\beta  \left( \mu(x)-m \phi(x) \right) \sum_{\varepsilon}n_{\varepsilon} -\beta \sum_{\varepsilon}n_{\varepsilon} \varepsilon \right),
\eqn
where $\mu(x)$ is the chemical potential at the volume,  $n_{\varepsilon}$ is the occupation number of $\varepsilon$ energy level,  $\beta$  is the inverse of the temperature of DM,  
and the gravitational potential is 
\bqn
\phi(r) = - 4\pi G \Big(\frac{1}{r} \int_0^r \rho(r') r'^2 dr' + \int_r^R \rho(r') r' dr' \Big). 
\eqn

Due to the absence of correlations between $\Delta V$ volumes in the vanilla collision-less model,  the probability of the state of the halo can be found by multiplying the probabilities of all $\Delta V$ volumes.  After a sum over all the possible halo states with their respective weights,  halo's partition function reads
\bqn
{\cal{Z}} &=&
\prod_{x}\Bigg(\sum_{N(x)} \exp\Big(\nb\\
&&\beta N\left(\mu-m\phi\right)-N\,\text{Ln}\left(\frac{\lambda^3\rho}{m}\right)+N
\Big)\Bigg),
\eqn
where $\lambda = \sqrt{(2 \pi m )^{-1} \, \beta}$,
the summation of $\varepsilon$ has been calculated,  and both the Stirling's approximation and $\rho = m N/\delta V$ have been utilized.  
In this equation,  we have used the Landau's approach of rearranging the summation over the halo states to only keep the summation over the parameter of interest.  
If we choose the infinitesimal volume to be arbitrarily small,  we can approximate the integral as $\int d^3x \simeq \sum_x \delta V$ and after defining $D\rho \equiv \prod_x \int d\rho(x)$,   halo's partition function can be expressed as follows
\bqn
{\cal{Z}} 
&=&\int D\rho \exp\Bigg(\nb\\
&&-\int d^3x\, \frac{\rho}{m}\Big[\text{Ln}\{ \frac{\rho}{m} \lambda^3\}-1
-\beta \mu + \beta m \phi 
\Big]\Bigg). \nb\\
\eqn

Therefore,  the effective free energy functional of DM mass density of the halo takes the following form \cite{2006PhyA..366..229L,2022Univ....8..386B}
\bqn
\lb{Eq:free-energy-ig}
\beta F\IG &=& \int d^3x \frac{\rho}{m} \Big[   \text{Ln}\{ \frac{\rho}{m} \lambda^3\} -1 -\beta \mu + \beta m \phi  \Big].
\eqn

A DM halo in a state of stability is positioned at the nadir of the effective free energy curve.  This premise implies the initial constraint on any proposed DM model of halos. Specifically, it requires that the first functional derivative of the system is zero at the halo's mean density
\bqn
\lb{Eq:FirstDerivEqual0}
\frac{\delta \beta F\IG}{\delta \rho (q)} \Big|_{\langle \rho \rangle } \simeq 0. 
\eqn
In an attempt to calculate the functional derivative, we make use of the following equations as expounded in appendix \ref{App:FirstFunctionalDerivativeOfPhi}
\bqn
\lb{Eq:some_functional_ders}
&&\frac{\delta}{\delta \rho(\vec{r}\,)} = \frac{\delta}{4 \pi r^2 \delta \rho(r)},\nb\\
&&\frac{ \delta \phi(\vec{r}_1) }{ \delta \rho(\vec{r}_2) } = -\frac{G}{r_{_>}},\nb\\
&&\int d^3x \,\rho(\vec{x}) \frac{\delta \phi(\vec{x})}{\delta \rho(\vec{x}\,') } = \phi(\vec{x}\,'),
\eqn
where $r_{_>}$ denotes the larger of $r_1$ and $r_2$ and we presume a spherical symmetry.

By applying $\frac{\delta}{\delta \rho (q)}$ to the right side of equation \eqref{Eq:free-energy-ig} and utilizing \eqref{Eq:some_functional_ders} to solve the integrals, the first functional derivative of the effective free-energy can be written as follows
\bqn
\frac{\delta \beta F\IG}{\delta \rho (q)} = \frac{4\pi q^2}{m}  \Big[   \text{Ln}\{ \frac{\rho}{m} \lambda^3\}  -\beta \mu + 2 \beta m \phi  \Big]. 
\eqn

The chemical potential $\mu(r)$, which remains unestablished by observations, can be tailored such that the first functional derivative of the effective free energy is null. Therefore, presuming that DM follows a simple system statistics and using the semi-equality mentioned above, we can express the chemical potential as follows
\bqn
\lb{Eq:mu_ig}
\mu\IG \simeq  \beta^{-1}\text{Ln}\{ \frac{\rho}{m} \lambda^3\} + 2 m \phi. 
\eqn

As has been previously demonstrated, the second functional derivative of either entropy or free energy is necessary to analyze the stability of gravitational systems \cite{1990PhR...188..285P, 2020EPJP..135..290C}. Given that our study reorganizes the sum in the partition function to define the effective free energy, the second functional derivative of the latter should analogously shed light on the stability of halos. Indeed, equation \eqref{Eq:mu_ig} represents a necessary but not sufficient condition for a stable halo. The rationale behind this is that the first functional derivative is also null at a maximum or extremum of an effective free-energy. In order for the halo to be at the minimum of the effective free-energy and thus be stable, the second functional derivative of the effective free-energy needs to be positive at any pair of arbitrary locations within the halo. However, the second functional derivative of the effective free-energy in equation \eqref{Eq:free-energy-ig} does not meet this requirement. It is expressed as follows
\bqn
\lb{Eq:2deriv_free-energy_ig}
\frac{\delta^2 \beta F\IG}{\delta \rho(\vec{r}_2)\delta \rho(\vec{r}_1)}  =
\frac{\delta(r_2-r_1)}{4\pi r_2^2 m \rho} - \frac{2\beta G}{r_{_>}}.
\eqn

It is evident that the second functional derivative of the simple cold collision-less DM model, as described in equation \eqref{Eq:2deriv_free-energy_ig}, is negative when $r_1 \neq r_2$. This implies that equation \eqref{Eq:mu_ig} corresponds to the maximum, rather than a minimum, of the effective free-energy as described in equation \eqref{Eq:free-energy-ig}. This suggests that the halo, while momentarily static, will eventually either condense towards a higher mass density profile or explode, thereby disappearing. Given that the effective free-energy equates to the negative logarithm of the probability of the mass density profile, the direction of evolution is random if a halo's chemical potential corresponds to equation \eqref{Eq:mu_ig}. We can demonstrate that if fluctuations cause the halo's chemical potential to exceed the value given in equation \eqref{Eq:mu_ig}, the halo's mass density escalates indefinitely towards higher values. Conversely, if the halo's chemical potential falls below the value given in equation \eqref{Eq:mu_ig}, the halo's mass density diminishes indefinitely towards lower mass density.

\section{Interactive Classic DM}
\lb{Sec:Interactive-DM}
This section aims to delve into the interactions between DM particles that may rectify the instability previously noted in the simple cold collision-less DM model. Specifically, we seek interactions that ensure the second functional derivative of the effective free-energy remains positive for any selected pairs of locations. 

In general, the Hamiltonian's inter-particle interactions can originate from either collective or fundamental forces. It can be represented as follows

\bqn
H_{I} = \frac{1}{2!}\sum_{ij} U_{_2ij} + \frac{1}{3!}\sum_{ijk} U_{_3ijk} + \cdots,
\eqn
where
the indices denote all the DM particles within the halo. Applying the identity $\sum_i = \int d^3x\,  \frac{\rho}{m}$, we can reformulate the equation in a continuum form
\bqn
H_I &=&  \frac{1}{2!}\int d^3x_1 d^3x_2  U_2(\vec{x}_1,\vec{x}_2) \rho(\vec{x}_1) \rho(\vec{x}_2) \nb\\
&+&  \frac{1}{3!}\int d^3x_1 d^3x_2 d^3x_3  U_3(\vec{x}_1,\vec{x}_2,\vec{x}_3) \rho(\vec{x}_1) \rho(\vec{x}_2)\rho(\vec{x}_3)\nb\\
&+& \cdots. 
\eqn
In this scenario, the partition function for the halo in the presence of classical interactions is given as
\bqn
{\cal{Z}} &=&\sum_{N}
\int d^{3N}q \exp\left(-\beta\left(\mu-m\phi\right)N -\beta H_{I}\right)
\nb\\
&& \times \int d^{3N} p  \exp\left( - \beta H\IG \right),
\eqn 

Here, the division is feasible because $H\IG$ depends solely on particle momentum, whereas $H_I$ and $\phi$ are functions of particle position. Given that $H_I$ can be entirely expressed in terms of $\rho(x)$ and considering that 
$$\rho(\vec{x})  \equiv m \sum_{i=1}^N\delta(\vec{x}-\vec{q}_i)$$ 
remains a function of the particle positions $\vec{q}_i$, the enumeration of energy states ($\sum_N \int d^{3N}q$) equates to that given by $\int D\rho$. 

Upon resummation, the partition function can be traditionally written as \cite{2006PhyA..366..229L}
\bqn
{\cal{Z}}  = \int D\,\rho \exp\Bigg(F\IG + F_{I}\Bigg). 
\eqn
The effective free energy here is partitioned into the simple term and the interaction term. In its most generic form, the latter can be expanded to
\bqn
\lb{Eq:free-energy-interaction}
&& F_I = \sum_{n=2}^{\infty}  \frac{1}{n!} \int \left(\prod_{a=1}^n d^3x_a \rho(\vec{x}_a)\right) 
 U_n(\vec{x}_1\cdots \vec{x}_n). 
\eqn

The modification of the
effective free-energy alters equation \eqref{Eq:FirstDerivEqual0}, which leads to the effective chemical potential, under the assumption of halo steadiness, being
\bqn
\mu \simeq \mu\IG + \frac{\delta \beta F_I}{\delta \rho(\vec{x})} \Big|_{\langle \rho \rangle}. 
\eqn

Additionally, the second derivative of the effective free-energy becomes
\bqn
\frac{\delta^2 \beta F}{\delta \rho(\vec{r}_2)\delta \rho(\vec{r}_1)}  &=&
\frac{\delta(r_2-r_1)}{4\pi r_2^2 m \rho} - \frac{2\beta G}{r_{_>}}\nb\\
&+&\beta U_2(\vec{x}_1,  \vec{x}_2) + \cdots.
\eqn

From the above, it's clear that by merely adjusting the two-body interaction, we can make the second functional derivative of the effective free-energy positive, irrespective of the chosen $\vec{x}_1$ and $\vec{x}_2$. This approach can potentially explain the long-term stability of halos. However, the two-body interaction, $U_2(\vec{x}_1,  \vec{x}_2)$, necessary for long-term stability must be non-local due to the existence of $2\beta G/r_{_>}$ and cnnot be proportional to the Dirac delta function.  This means that the interaction should remain positive and non-zero when the interacting particles' positions vary. Given that gravity, already accounted for, is the only known force capable of operating over galactic distances, $U_2(\vec{x}_1,  \vec{x}_2)$ must either have an unusual nature, such as emerging effectively from quantum effects, or be collectively produced by other phenomena like baryonic feedback. 

Since most baryons in galaxies are located at the center, an intriguing research direction could involve exploring whether baryonic feedback can generate an effective two-body interaction that remains non-zero even between two points that are both distanced from the center.

\section{Quantum DM}
\lb{Sec:Quantum-DM}
Considering DM models with appreciable quantum effects, we examine the alterations to the partition function of a non-interacting DM system. The formal partition function of such a system is expressed as
\bqn
{\cal{Z}} = \int d^3x_1 \cdots d^3x_N \sum_{E,N} e^{-\beta \left(E-\mu N \right)} 
|\Psi_E(\vec{x}_1\cdots \vec{x}_N)|^2,\nb\\
\eqn
with $\Psi$ symbolizing the quantum state of the halo, which is detailed as
\bqn
\Psi_E(\vec{x}_1\cdots \vec{x}_N) = \left( N! \right)^{-\frac{1}{2}}
\sum_p p\left[u_{\epsilon_1}(\vec{x}_1) \cdots u_{\epsilon_N}(\vec{x}_N)\right].\nb\\
\eqn
Here, $p$ signifies the permutation operator with $\sum_p$ pointing to all possible permutations, and $u_{\epsilon_i}(\vec{x}_i)$ defines the wave function of the $i^{\text{th}}$ particle satisfying the Schrodinger equation. 

In scenarios where DM particles are non-interacting, the total energy of the halo equates to the aggregation of the energies of the individual particles, represented as $E = \sum_{i=1}^N \epsilon_i$. Consequently, the partition function can be rewritten as 
\bqn
{\cal{Z}} &=& \int d^3x_1 \cdots d^3x_N \sum_{N} e^{\beta \mu N } \frac{1}{N!} \sum_p 
	\Bigg[\nb\\
	&&\left(
			\sum_{\epsilon_1} u^*_{\epsilon_1}(p\vec{r}_1)u_{\epsilon_1}(\vec{r}_1)e^{-\beta \epsilon_1}
		\right) \nb\\
		&&~~\times \cdots \times \nb\\
	&&\left( 
		 	\sum_{\epsilon_N} u^*_{\epsilon_N}(p\vec{r}_N)u_{\epsilon_N}(\vec{r}_N)e^{-\beta \epsilon_N}
		 \right)
	\Bigg].
\eqn

After introducing
\bqn
f\left(p\vec{r}_i, \, \vec{r}_i | \beta\right)
	\equiv 
		\frac{\lambda^3}{V} \sum_{\epsilon} u^*_{\epsilon}(p \vec{r}_i) u_{\epsilon}(\vec{r}_i) e^{-\beta \epsilon},
\eqn
and applying the properties of natural logarithm and Stirling's approximation of the logarithm of $N!$, the partition function assumes the following form
\bqn
{\cal{Z}} &=& \int d^3x_1 \cdots d^3x_N \sum_{N} e^{\beta \mu N }
\exp\Bigg(\nb\\
	&&-N \text{Ln}\left(\frac{\lambda^3 N}{V}\right) + N + \nb\\
	&&\text{Ln}\Bigg(
							\sum_p \Big[ f\left(p\vec{r}_1,  \vec{r}_1 | \beta\right) \cdots f\left(p\vec{r}_N,  \vec{r}_N | \beta\right)\Big]
						\Bigg).
	 \nb\\
&&\Bigg).
\eqn

Using the demonstration in \cite{2022Univ....8..386B},  the resultant partition function can be expressed as a combination of the typical cold collision-less DM equation \eqref{Eq:free-energy-ig} and a corrective term due to quantum effects:
\bqn
\lb{Eq:QuantumEffectiveFreeEnergy_final}
&&F =  F\IG + F_{\text{Q.M.}}\nb\\
&& \beta F_{\text{Q.M.}} = 
-\beta \text{Ln}\Bigg(
							\sum_p \Big[ f\left(p\vec{r}_1,\vec{r}_1 | \beta\right) \cdots f\left(p\vec{r}_N,  \vec{r}_N | \beta\right)\Big]
						\Bigg). \nb\\
\eqn
Upon evaluating this effective free-energy and comparing it with equation \eqref{Eq:free-energy-interaction}, it becomes clear that, in this context, a quantum description of DM is equivalent to a classical model of DM that includes specific types of interactions. A similar equivalence of statistical models for, non-gravitational, quantum and classic interactive systems was introduced by Uhlenbeck and Gropper in the 1930s \cite{PhysRev.41.79}.

\subsection*{When $f \left(\vec{r}, \vec{r}\right) = 1$ \& $f \left(\vec{r}_1, \vec{r}_2\right) \ll 1$ for large $|\vec{r}_1 - \vec{r}_2|$:}
In this subsection, we consider a DM halo that exhibits weak entanglement. In this case, the function $f \left(\vec{r}_1, \vec{r}_2\right)$, which characterizes the degree of quantum effects, only holds significant value when the locations $\vec{r}_1$ and $\vec{r}_2$ are close.  Given this scenario, we can express the quantum modification to the effective free-energy, as depicted in equation \eqref{Eq:QuantumEffectiveFreeEnergy_final}, in an expanded form
\bqn
F_{\text{Q.M.}} &=& -\text{Ln}\Bigg( 1 \pm \frac{1}{2} \sum_{ij} |f \left(\vec{r}_i, \vec{r}_j\right)|^2 + \dots \Bigg),\nb\\
&\simeq & \mp  \frac{1}{2} \sum_{ij} |f \left(\vec{r}_i, \vec{r}_j\right)|^2,
\eqn
where each term inside the logarithm represents the number of permutations,  and we have neglected higher order terms. 
Also,  in terms with dual sign, the upper and lower signs correspond to bosons and fermions respectively. 
Upon employing the relation $\sum_i = \int d^3x \rho/m$, the above equation can be transformed as
\bqn
F_{\text{Q.M.}} =  \frac{ \mp 1}{2m^2}\int d^3x_1 d^3x_2 |f \left(\vec{x}_1, \vec{x}_2\right)|^2 \rho(\vec{x}_1)\rho(\vec{x}_2). \nb\\
\eqn

Consequently, the additional term in the second derivative of the effective free-energy assumes the form 
\bqn
\lb{Eq:d2F_QM_FermionBoson}
\frac{\delta^2 \beta F}{\delta \rho(\vec{r}_2)\delta \rho(\vec{r}_1)}  =
\frac{\delta(\vec{r}_2-\vec{r}_1)}{m \rho} - \frac{2\beta G}{r_{_>}} \mp \frac{1}{m^2} |f \left(\vec{x}_1, \vec{x}_2\right)|^2. \nb\\
\eqn

From the above discussion, we infer that in a DM halo where quantum effects are not strong, a bosonic DM tends to exacerbate the instability problem by rendering the second derivative of the effective free energy more negative. On the other hand, a fermionic DM has the potential to alleviate the instability for closely situated location pairs. Nevertheless, in the event that $\vec{x}_1$ and $\vec{x}_2$ are considerably distant, the corrective term loses significance, and the halo reverts to an unstable state. 

By employing equation \eqref{Eq:d2F_QM_FermionBoson}, we can deduce that halos composed of bosons, where quantum effects are substantial, exhibit sharply defined edges. In contrast, halos consisting of fermions with considerable quantum influences feature edges that are less sharply delineated and extend more diffusely toward infinity.
The underlying reason for this behavior is that as we venture further into the halo's outer regions, we eventually reach a distance where quantum effects can be adequately addressed using perturbation methods. Given that the value of \( r_{_>} \) is relatively large at these distances, the influence of gravitational instability becomes minimal. Nonetheless, the instability inherent to bosons, as indicated in equation \eqref{Eq:d2F_QM_FermionBoson}, persists. This confines the bosonic halo to regions where quantum effects cannot be treated perturbatively and sharply cuts the outer region.

\section{Most General Model of DM Perturbations}
\lb{Sec:General-DM-perturbation}
The enduring mystery surrounding the nature of DM does not preclude us from leveraging the observed (at least quasi-)stability of DM halos to our advantage.  For one, stability constraints imply that when we expand the effective free-energy of any given DM model around a mass profile $\rho_{_O}$, as established by experiments,  only the leading terms significantly contribute. Furthermore, these expansion coefficients are determinable based on the symmetries of the density perturbations around $\rho_{_O}$ in the halo. 

Drawing parallel to other statistical systems \cite{Wu2021}, it can be suggested that the symmetry of these fluctuations, not the specific principles of the DM model, dictates the effective free-energy of mass density perturbations in the halo. Consequently, a broad and seemingly disparate collection of DM models may actually fall under the same universality class, and hence, predict similar fluctuations around the chosen background mass density $\rho_{_O}$.

To gain deeper insights, we undertake an expansion of the most general effective free-energy of a DM halo around $\rho_{_O}$, as given by
\bqn
\lb{Eq:F_varphi_most_general}
\beta F[\varphi] = \sum_{n=1}^{\infty} \int d^3x_1 \cdots d^3x_n \, c^{(n)} \varphi_1 \cdots \varphi_n,
\eqn
where
\bqn
\lb{Eq:coef_varphi_most_general}
&&c^{(n)}(\vec{x}_1,\cdots, \vec{x}_n)   \equiv \frac{\delta^n \beta F}{\delta \rho(\vec{x}_1) \cdots \rho(\vec{x}_n)} \Big|_{\rho _{_O}} 
 \rho_{_O}(\vec{x}_1)  \cdots  \rho_{_O}(\vec{x}_n) ,\nb\\
~\nb\\
&&\varphi_i \equiv \frac{ \rho(\vec{x}_i) -  \rho_{_O}(\vec{x}_i)}{ \rho_{_O}(\vec{x}_i)},
\eqn
where a constant is absorbed by the  normalization factor.  It's important to note that the first functional derivative of the effective free-energy at $\rho_{_O}$ isn't set to zero given that the choice of $\rho_{_O}$ isn't necessarily equal to the average mass density $\langle \rho \rangle$, and retains a degree of arbitrariness.

Having established this, we can proceed to study potential predictions of DM models for mass density fluctuations $\varphi(x)$ by systematically varying the $c^{(n)}$ coefficients. This essentially amounts to transitioning from one universality class to another. Observational data can then be used to evaluate these classes of DM models based on the predictions they make. For instance, the n-body correlation between mass densities, i.e. $\langle \varphi(\vec{x}_1) \cdots \varphi(\vec{x}_n)\rangle$, across the halo is directly linked to the selection of $c^{(n)}$ and can be tested using observations. 

In our previous work, we demonstrated how two-body correlations $\langle \varphi  (\vec{x}_1) \varphi(\vec{x}_2)\rangle$, as derived from observations, can help in refining the parameter space of DM models \cite{2022Univ....8..386B}. Continuing along this vein, we show here how the average $\langle \varphi  (\vec{x
}) \rangle$ - constructible in a similar manner from observational data - along with stability constraints can aid in constraining the parameters of DM models.

\subsection*{A DM Model with a Two-Body Interaction: A Showcase}
In the present study, we strive to illustrate a straightforward extension to the classic model of cold, collision-less DM. This extension incorporates two-body interactions into the effective free-energy of the dark matter halo
\bqn
\lb{Eq:ShowCase_G}
&&F =  F\IG +  \frac{1}{2} \int d^3x_1 d^3x_2\, \rho(\vec{x}_1)\rho(\vec{x}_2) U_2(\vec{x}_1,  \vec{x}_2). 
\eqn

Once we have the complete form of $F$, equation \eqref{Eq:coef_varphi_most_general} can be utilized to compute the coefficients that describe the mass density fluctuation effective free-energy represented in equation \eqref{Eq:F_varphi_most_general}. The coefficients are given by the following expressions
\bqn
&&c^{(1)} = \frac{\rho_{_O}(\vec{x})}{m}\Bigg(\text{Ln}\left( \frac{\rho_{_O}(\vec{x})}{m} \lambda^3\right)  -\beta \mu(\vec{x}) + 2 \beta m \phi  \Bigg)\nb\\
&&~ + \int d^3x'  U_2(\vec{x},  \vec{x}\,') \rho_{_O}(\vec{x'}) \rho_{_O}(\vec{x}),\nb\\
&&c^{(2)}  = \Bigg( \frac{\delta(\vec{x}_2-\vec{x}_1)}{ m \rho_{_O}(\vec{x}_1)} - \frac{2\beta G}{r_{_>}}
+  \beta U_2(\vec{x}_1,  \vec{x}_2) \Bigg) \rho_{_O}(\vec{x}_1) \rho_{_O}(\vec{x}_2) ,\nb\\
&&c^{(3)} = -\frac{1}{m}\delta(\vec{x}_2-\vec{x}_1) \delta(\vec{x}_3-\vec{x}_2)\rho_{_O}(\vec{x}_3),\nb\\
&&c^{(4)} = \frac{2}{m}\delta(\vec{x}_2-\vec{x}_1) \delta(\vec{x}_3-\vec{x}_2)
\delta(\vec{x}_4-\vec{x}_3)\rho_{_O}(\vec{x}_4).
\eqn

A comprehensive analysis of possible $U_2(\vec{x}_1,  \vec{x}_2)$ interaction terms and their observational implications is beyond the scope of the present work. In this paper, we simplify the showcase by selecting the interaction such that $- \frac{2\beta G}{r_{_>}} +  \beta U_2(\vec{x}_1,  \vec{x}_2) \simeq 0$.
It seems unlikely that this interaction originates from a fundamental force. Instead, it may be more plausible to consider that the interaction is mediated by phenomena that become significant towards the center of a halo. Regardless of its origin, we are using this interaction as a toy model for the purpose of this presentation, offering a simplified way to explore and understand the system in question.
Consequently, the effective free-energy of our chosen showcase model is
\bqn
\lb{Eq:ShowCase_F_V2}
&&\beta F = \int d^3x\,\frac{\rho_{_O}}{m}(\vec{x}) g[\varphi],\nb\\
&&g[\varphi] \equiv h(x) \varphi(x) + \varphi^2(x) - \varphi^3(x) + 2 \varphi^4(x), \nb\\
&&h(\vec{x})
 \equiv  \frac{m}{\rho_{_O}(\vec{x})} c^{(1)},\nb\\
\eqn
where the term $h(\vec{x})$ is dependent on the specific halo under investigation as the chemical potential $\mu$ is influenced by the halo's environment and other characteristics. To emphasize the potency of the statistical field theory approach, we make the assumption that the chemical potential and the temperature of a halo are determined through observations and that these observations suggest $h(r) = r^{-3}$. 
In other words, at the moment, direct observational measurements of the chemical potential and the temperature remains a challenge. Therefore, we have to make certain assumptions to account for this limitation. To that end, we have translated our assumptions about the chemical potential and the temperature into a specific form for the \( h(r) \) term. 
With this assumption, we can plot the functional $g[\varphi]$ as a function of the mass density fluctuations $\varphi$, as shown in figure \ref{fig:f_phi}. 
For the sake of argument,  we would like to see how the mean of the mass density fluctuations can be utilized to evaluate the model defined in equation \eqref{Eq:ShowCase_G}. 

In figure \ref{fig:f_phi}, assuming that the radius of the halo is scaled to one, the curve with label $r=1$ belongs to the edge of the halo. It shows that at this distance from the center, the minimum of $g[\varphi]$ is at $\varphi =0$ and the average of the fluctuations tend to be zero. The curve representing distance of 0.9 has a minimum slightly less than zero. This means that if we measure the fluctuations at that distance, the average would have a net negative value. As is evident from the rest of the curves, this general trend exist that as we move toward the center, the average of the perturbations deviate further away from zero, and the width of the well shape of the curves increases as the distance decreases toward the center. Finally, at a distance equal to 10\% of the halo's radius, the curve labeled with $r=0.1$ shows no minimum, i.e. the width of the well has become substantially large, indicating the instability of the toy model's halo in that region. In other words, from figure \ref{fig:f_phi}, we note that the minimum of the functional $g[\varphi]$,  representing where $\langle \varphi \rangle$ is located,  shifts from a significantly negative value at the halo's center to zero in the outer region of the halo. Therefore, by measuring the DM mass density as a function of distance from the center of a halo and subtracting it from the selected mass profile,  for example NFW or Burkert, we can construct a phenomenological $\langle \varphi \rangle$ that can be used to test the predictions of a given class of DM models,  consequently restricting the parameter space of all DM models belonging to that class.

\begin{figure}[h]
  \centering
  \includegraphics[width=\columnwidth]{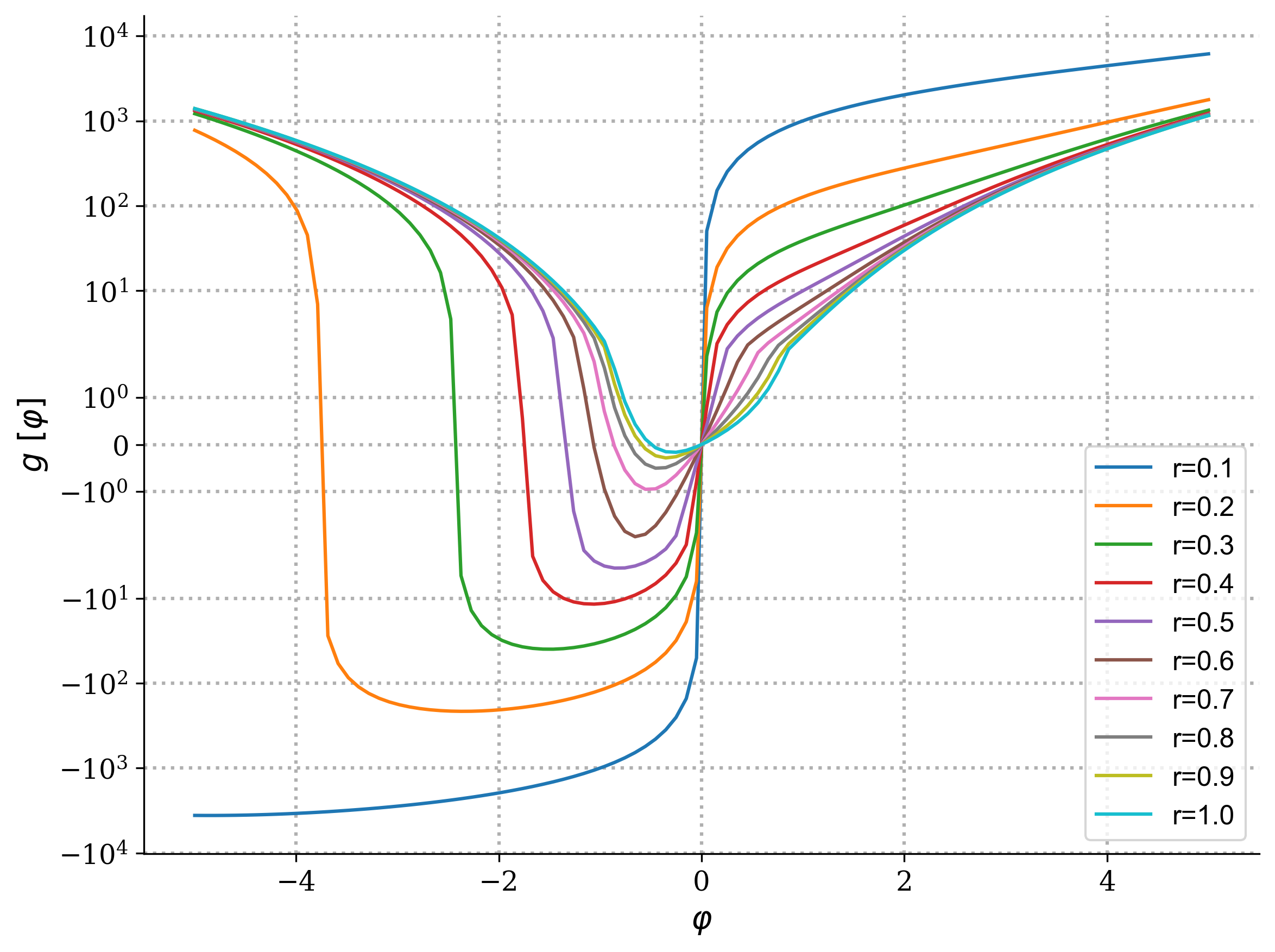}
  \caption{
	The integrand of the effective free energy in equation \eqref{Eq:ShowCase_F_V2} in term of the mass density fluctuations of DM at various distances $r$ from the center of DM halo.  As can be seen in this figure,  the most probable value of the mass density fluctuations shifts from negative values to zero for locations at the center and edge of the halo respectively.  Also, since $r=0.1$ curve has no minimum, the plot indicates that the model represented by equation \eqref{Eq:ShowCase_F_V2} is unstable at the center of the halo and will tend to have less DM in the future. Exploring whether the observed instability at the center propagates to destabilize the entire halo would be a compelling avenue for future research. To address this question, the static approaches outlined in this paper would need to be expanded upon. Specifically, a dynamic analysis of the halo's behavior is required, placing this inquiry firmly within the scope of existing N-body simulations.}
\lb{fig:f_phi}
\end{figure}

\section{Conclusion} 
\lb{Sec:Conclusion}
We presented a robust methodology to investigate the stability of DM halos. Our emphasis was on the critical role of non-local inter-particle interactions, be they fundamental or effective, in sustaining the stability of these halos. We underscored the shortfalls of conventional cold collision-less DM models in predicting stable halos without taking into account a ``non-local'' interaction within the effective free energy of the halo. These interactions might stem from elements like baryonic feedback, self-interactions, or the inherent quantum features of dark particles. Stability, as we concluded, required substantial effective interactions between any two points inside the halo, independent of their distance from the center.

The methodology we proposed serves as a systematic framework for scrutinizing classes of DM models and for refining their parameter spaces. Based on our investigation, we inferred that DM halos, where the divergence from the standard cold collision-less framework was restricted to areas near the halo center, were not expected to maintain stability in their outer regions.

We showed that the problem of instability within DM halos could not be sufficiently resolved by resorting to perturbative quantum effects. This problem was not as severe for fermionic dark matter, yet it was considerably more pronounced in the case of bosonic DM.

In showcasing the potential of this cross-model approach, we delved into the most encompassing form of the effective free-energy around a selected mass profile. We used a model in which the deviation from the standard cold collision-less DM model was characterized by a two-body interaction within the effective free-energy. We demonstrated how to utilize observational data to examine different classes of DM models.

\bibliography{Refs}

\appendix
\section{First functional derivative}
\lb{App:FirstFunctionalDerivativeOfPhi}

To derive the functional derivatives in this article,  we start with definition of the derivative of a functional $B[\rho(\vec{x})]$ with respect to $\rho(\vec{x})$
\bqn
\lb{Eq:App1}
\frac{\delta B[\rho(\vec{x})]}{\delta \rho(\vec{x}^{\prime})} \equiv \lim _{\varepsilon \rightarrow 0} 
\frac{B[\rho(\vec{x})+\varepsilon \delta(\vec{x}-\vec{x}^{\prime})]-B[\rho(\vec{x})]}{\varepsilon},\nb\\
\eqn

Let's now define the following functional
\bqn
B[\rho] \equiv \int d^3x\, \rho(\vec{x}). 
\eqn
Through a simple substitution,  we can see that 
\bqn
B[\rho(\vec{x}) + \varepsilon \delta(\vec{x}-\vec{x}^{\prime})] = B[\rho(\vec{x})] + \varepsilon. 
\eqn
Putting this back into equation \eqref{Eq:App1},  we can conclude that 
\bqn
&&\frac{\delta \rho(\vec{x}^{\prime})}{\delta \rho(\vec{x})} 
 = \delta (\vec{x} - \vec{x}^{\prime}),
\nb\\
&&\frac{\delta \rho(r^{\prime})}{\delta \rho(r)} 
 = \delta (r - r^{\prime}). 
\eqn
Using this equation and since 
$$
\frac{\delta }{\delta \rho(r)}  \int dr \, \rho 
= 
\frac{\delta }{\delta \rho(\vec{x})} \int d^3x \, \rho =1,
$$
and using the spherical symmetry we can conclude that 
\bqn
\lb{App_Eq:d_drho}
\frac{\delta}{\delta \rho(\vec{r}\,)} = \frac{\delta}{4 \pi r^2 \delta \rho(r)}.
\eqn
~\\

To show that $\frac{ \delta \phi(\vec{r}_1) }{ \delta \rho(\vec{r}_2) } = -\frac{G}{r_{_>}}$,  we directly apply 
$\frac{\delta }{\delta \rho(r)} $  to the definition of the gravitational potential 
\bqn
&&\frac{\delta \phi(r)}{\delta \rho \left(r^{\prime \prime}\right)}=\nb\\
&&-4 \pi G\left(\frac{1}{r} \int_0^r \delta\left(r^{\prime}-r^{\prime \prime}\right) r'^{2} d r^{\prime}+\int_r^R \delta\left(r^{\prime}-r^{\prime \prime}\right) r^{\prime} 
d r^{\prime}\right). \nb\\
\eqn

After working out the integrals,  it is easy to see that 
\bqn
\lb{App_eq:dphi_drho}
\frac{\delta \phi(r)}{\delta \rho \left(r^{\prime}\right)} 
= 
-4 \pi G \begin{cases}
 \frac{r^{\prime 2}}{r} & r^{\prime}<r \\ 
 ~\\
 r^{\prime } & r^{\prime }>r
\end{cases} 
\eqn
This equation can be easily converted to the format we presented above after using equation \eqref{App_Eq:d_drho}.

Finally,  after substituting equation \eqref{App_eq:dphi_drho} into 
$$
\int d^3x \,\rho(\vec{x}) \frac{\delta \phi(\vec{x})}{\delta \rho(\vec{x}\,') }
$$ 
and after working out the integration,  we can find that it is equal to $\phi(\vec{x}\,  ')$.

\end{document}